%% file: main.tex
\documentclass[twocolumn,aps,prl,superscriptaddress]{revtex4-2}  
\usepackage{graphicx}  
\usepackage{bm}        
\usepackage{amssymb}   
\usepackage{xcolor}
\usepackage{amsmath}

\usepackage[textheight=672pt,textwidth=510pt]{geometry}

\usepackage{layout}

\definecolor{linkColor}{RGB}{0,80,150}
\usepackage{hyperref}
\hypersetup{
    colorlinks=true,
    allcolors=linkColor,
    pdfborder={0 0 0},
    pdfencoding = auto
}

\usepackage[normalem]{ulem}

\DeclareMathOperator{\Tr}{Tr}

\newcommand{\evec}{\mathbf{e}}

\begin{document}

\title{Patterning of morphogenetic anisotropy fields}

\author{Zihang Wang}
\author{M.\ Cristina Marchetti}
\affiliation{Department of Physics, University of California Santa Barbara, Santa Barbara, California 93106, USA}
\author{Fridtjof Brauns}
\email{fbrauns@ucsb.edu}
\affiliation{Department of Physics, University of California Santa Barbara, Santa Barbara, California 93106, USA}
\affiliation{Kavli Institute for Theoretical Physics, University of California Santa Barbara, Santa Barbara, California 93106, USA}
 
\begin{abstract}
Orientational order, encoded in anisotropic fields, plays an important role during the development of an organism.
A striking example of this is the freshwater polyp \textit{Hydra}, where topological defects in the muscle fiber orientation have been shown to localize to key features of the body plan. 
This body plan is organized by morphogen concentration gradients, raising the question how muscle fiber orientation, morphogen gradients and body shape interact.
Here, we introduce a minimal model that couples nematic orientational order to the gradient of a morphogen field. We show that on a planar surface alignment to a radial concentration gradient can induce unbinding of topological defects, as observed during budding and tentacle formation in \textit{Hydra}, and stabilize aster/vortex-like defects, as observed at a \textit{Hydra}'s mouth. 
On curved surfaces mimicking the morphologies of \textit{Hydra} in various stages of development---from spheroid to adult---our model reproduces the experimentally observed reorganization of orientational order. 
Our results suggest how gradient alignment and curvature effects may work together to control orientational order during development and lays the foundations for future modeling efforts that will include the tissue mechanics that drive shape deformations.
\end{abstract}

\maketitle

\definecolor{lightblue}{RGB}{200,220,255}
\setlength{\fboxsep}{0.03\linewidth}
\noindent\fcolorbox{white}{lightblue}{\parbox{0.94\linewidth}{%
\textbf{\sffamily Significance statement.}
Life has brought about a remarkable diversity of organism shapes. During development, an organism is sculpted by mechanical forces that are generated within its tissues and directed by orientational patterns, as found in the orientation of muscle fibers. How orientational order is coupled to the morphogen concentration fields that organize the body plan is not well understood. Using a minimal model for how orientational patterns couple to gradients of morphogen expression, we examine how they jointly organize the shape of developing organisms. The model recapitulates the reorganization of muscle fiber orientation during key developmental stages of the freshwater polyp \textit{Hydra}.
Our results lay a foundation for disentangling the interplay between tissue mechanics, orientational order, and morphogen gradients.
}}
\vspace{0.5em}

Morphogenesis, the process by which an organism acquires its shape, is evidently a mechanical process. The physical quantities that describe it---such as displacements, stresses and strains---are vectors and tensors which carry orientational information. 
The role of such orientational order, or anisotropy, in biological systems has received substantial attention in recent years. Examples include planar cell polarity \cite{Eaton.Julicher2011, Devenport2014, Dye.etal2021}, directed auxin transport in plants \cite{Heisler.etal2010}, anisotropic myosin-generated stresses driving tissue elongation \cite{Streichan.etal2018,Lefebvre.etal2022,Stokkermans.etal2022,Mitchell.etal2022}, and orientational order in cell monolayers \cite{Kawaguchi.etal2017,Guillamat.etal2022,Turiv.etal2020}.
A particularly striking example is the \emph{nematic} organization of supracellular actin fibers (known as myonemes) in the ectoderm and endoderm of \textit{Hydra} \cite{Aufschnaiter.etal2017}. In the ectoderm the myonemes are aligned along the the body axis, while the endodermal actin fibers are aligned perpendicular to those in the ectoderm, i.e.\ azimuthal to the cylindcrical body. Topological defects in the fiber orientation are located at key parts of the body plan (head, foot, bases and tips of tentacles) \cite{Maroudas-Sacks.etal2021}. 
This has sparked growing interest in nematics on curved and deforming surfaces \cite{Kralj.etal2011, Hoffmann.etal2022b, Vafa.Mahadevan2022, Nestler.etal2018}.
On the other hand, \textit{Hydra's} body plan is known to be organized by the concentration profiles of specific proteins called morphogens \cite{Turing1952, Wolpert1969, Bottger.Hassel2012}.
Furthermore, experiments indicate that actin fibers in \textit{Hydra} might align along gradients in a central player of the Wnt-morphogen pathway that sets Hydra's up the head-to-foot body axis \cite{Wang.etal2020, Shani-Zerbib.etal2022}.
While the molecular mechanism underlying this coupling remains to be investigated, these experiments suggests that alignment of myonemes along morphogen gradients may drive the reorganization of their orientation during key morphogenetic processes, such as body-plan development and budding in \textit{Hydra} \cite{Otto1977,Philipp.etal2009,Aufschnaiter.etal2017,Maroudas-Sacks.etal2021,Shani-Zerbib.etal2022} and other cnidaria \cite{Fritz.etal2013,Sinigaglia.etal2020}.

\begin{figure*}
    \centering
    \includegraphics{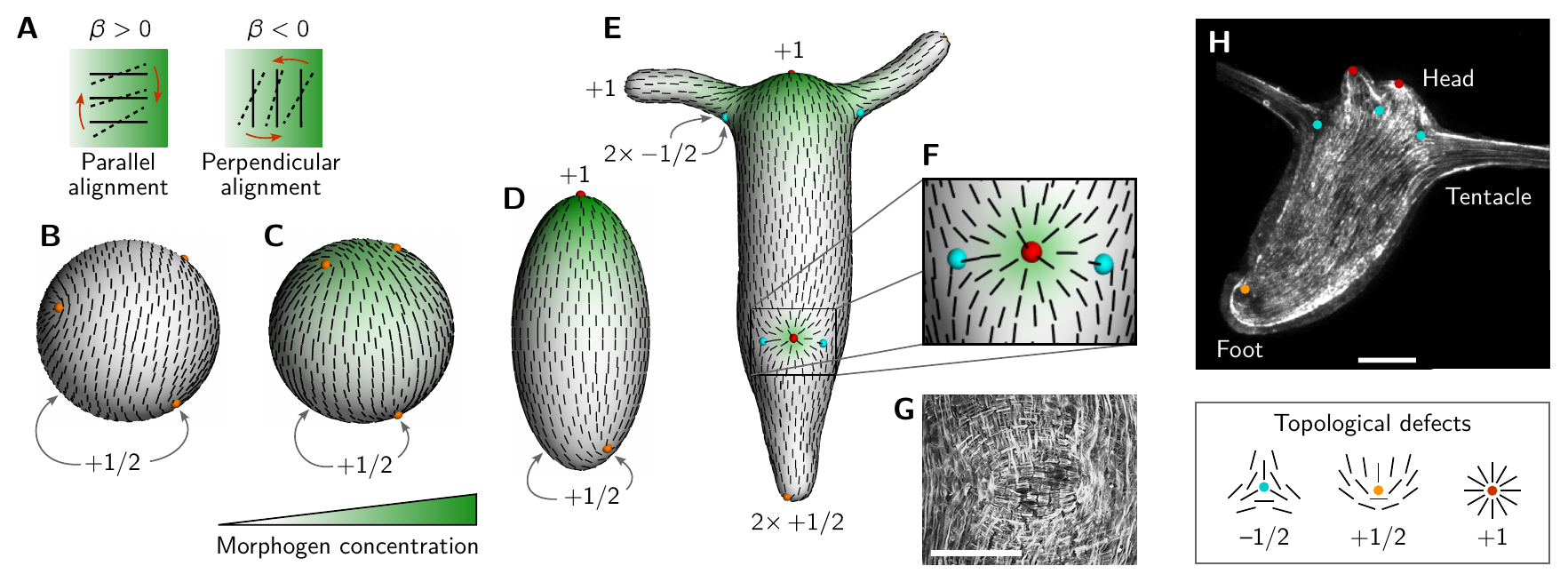}
    \caption{A minimal model coupling nematic orientational  to a morphogen gradient on curved surfaces recapitulates the salient features of actin fiber organization during \textit{Hydra} regeneration.
    \textbf{A} The directors (myonemes) align parallel or perpendicular to the concentration gradient, depending on the sign of the coupling constant $\beta$ in Eq.~\eqref{eq:coupling}.
    \textbf{B} On a sphere, a total defect charge of $+2$ is required by topology. In the absence of gradient alignment, four $+\frac12$ defects (orange points) arrange in a tetrahedral configuration.
    \textbf{C} Alignment to a morphogen gradient drives two $+\frac12$ defects towards the concentration maximum. 
    \textbf{D} After elongation of the body into an ellipsoid, the $+\frac12$ defects are attracted to regions of positive Gaussian curvature. For sufficiently high curvature and/or gradient alignment two $+\frac12$ defects merge into a $+1$ defect. 
    \textbf{E} On a surface modeled after the typical morphology of adult \textit{Hydra}, topological defects localize at the positions observed \textit{in vivo} by the interplay of the head-organizer morphogen gradient (green shading) and intrinsic curvature (base and tips of tentacles, foot). Note that each tentacle has a +1 defect at its tip and two $-\frac12$ defects at its base; cf.\ Movie~1.
    \textbf{F} Defect unbinding due to gradient alignment at an incipient bud.
    \textbf{G} Phalloidin-GFP stained actin fibers in the ectoderm (vertical orientation) and endoderm (horizontal) at an incipient bud (adapted from \cite{Philipp.etal2009}). 
    \textbf{H} Actin fibers in the ectoderm of an adult \textit{Hydra} (adapted from \cite{Maroudas-Sacks.etal2021}).
    (Scale bars 100$\,\mu$m.)
    }
    \label{fig:overview}
\end{figure*}

Here, we examine the relative role of morphogen gradients and body shape/topology in controlling a nematic texture. To this end, we introduce a minimal model for alignment of a nematic texture, representing the orientational order, to a prescribed morphogen concentration profile on surfaces of specified shape.
Figures~\ref{fig:overview}B--E show the steady state configurations predicted by this model on prescribed surfaces mimicking the sequence of morphologies of regenerating \textit{Hydra}.
These configurations reproduce the salient features of the reorganization of the myoneme orientational order during \textit{Hydra} regeneration.

\textit{Hydra} fully regenerate from small excised tissue fragments which first close into a spheroid \cite{Bode.Bode1980,Bode.Bode1984}. Nematic order of actin fibers is quickly reestablished, giving rise to four $+\frac12$ defects (Fig.~\ref{fig:overview}B), while additional pairs of $\pm\frac12$ defects quickly annihilate \cite{Maroudas-Sacks.etal2021}.
Subsequently, two of these defects migrate towards each other and eventually merge into a $+1$ defect at a location that coincides with the future formation of the head of the animal, suggesting that the body axis is already established at this stage (Fig.~\ref{fig:overview}C). This body axis is determined by the ``head organizer'' (Wnt-pathway) morphogen gradient (green shading), which is partially inherited from the tissue fragment's parent animal \cite{Shani-Zerbib.etal2022,Bode2009}.
Following the establishment of the body axis, the spheroid elongates into a prolate ellipsoidal shape (Fig.~\ref{fig:overview}D) and finally attains the morphology of the adult \textit{Hydra} (Fig.~\ref{fig:overview}E,H). At these later stages, the two remaining $+\frac12$ defects move towards the foot (the pole opposite to the head) and eventually merge there.

Topological defects serve as organizational centers for orientational order, making them key to understand orientational patterning.
While the initial phases of regeneration only involve migration and mergers of existing topological defects, budding and tentacle formation require unbinding of defects in a previously defect-free region (see Fig.~\ref{fig:overview}F,G). In the following we demonstrate how both the major reorganization of the myoneme orientational order during regeneration and the budding/tentacle formation can be attributed to the interplay between fiber alignment to gradients of a ``head organizer'' morphogen and the influence of surface curvature.  

To disentangle the interplay between gradient alignment and curvature effects, we first examine how gradient alignment on a planar surface can induce (\textit{i}) unbinding of $\pm\frac12$ defect pairs from a defect-free background and (\textit{ii}) the merger of two $+\frac12$ defects into a +1 defect.
We then turn our attention to simple curved surfaces, spheres and ellipsoids, to systematically study the interplay between topology, geometry and gradient alignment.
Finally, we discuss how this interplay can reproduce the key features of myoneme reorganization as observed in regenerating \textit{Hydra} and argue that gradient alignment of myonemes plays a key role in this process.

\section{\sffamily Results}

\textbf{\sffamily Planar system.}
The myonemes in \textit{Hydra}'s ectoderm exhibit strong orientational order \cite{Kass-Simon1976, Otto1977, Philipp.etal2009,Aufschnaiter.etal2017,Maroudas-Sacks.etal2021}, suggesting that neighboring myonemes align to one another. In our minimal model, we describe this orientational order on a coarse-grained level using the nematic tensor $Q_{ij}=S (n_in_j-\delta_{ij}/2)$ which represents the local director orientation, $\mathbf{n}$, and the local degree of order $S$. Microscopically, we can think of the director field as representing the local orientations of individual myonemes. On a coarse grained level, the director indicates the mesoscopic average of myoneme orientations and $S$ indicates how well myonemes are aligned locally.
We describe the dynamics of the director field in terms of a free energy functional $E = \int \mathrm{d}x^2 (f_\mathrm{LdG} + f_\mathrm{a})$ whose minima are the steady state configurations of the system.
The first contribution to the free energy density, $f_\mathrm{LdG}$, is the Landau–de-Gennes free energy density describing local alignment of the directors:
\begin{equation} \label{eq:free-energy-plane}
    f_\mathrm{LdG} = \frac{1}{2} (\Tr\mathbf{Q}^2-1) \Tr\mathbf{Q}^2 + \frac{K}{2}(\partial_i Q_{jk})^2,
\end{equation}
The first term drives the scalar order parameter towards $S = 1$ and the second term penalizes bend and splay deformations of the director field with the single Frank elastic constant $K$ (one-constant approximation). 
The second contribution to the free energy density, $f_\mathrm{a}$, captures the tendency of the director to align with the concentration gradient.
The simplest possible such coupling is
\begin{equation} \label{eq:coupling}
    f_\mathrm{a} = -\beta \, Q_{ij}(\nabla_{\!i} c)(\nabla_{\!j} c) = -\beta S \, (\mathbf{n} \cdot \nabla c)^2 
\end{equation}
where $\beta$ denotes the strength of the alignment.
For $\beta > 0$ ($\beta < 0$) this favors parallel (perpendicular) alignment of the director to the gradient (see Fig.~\ref{fig:overview}A).

\begin{figure*}
    \centering
    \includegraphics{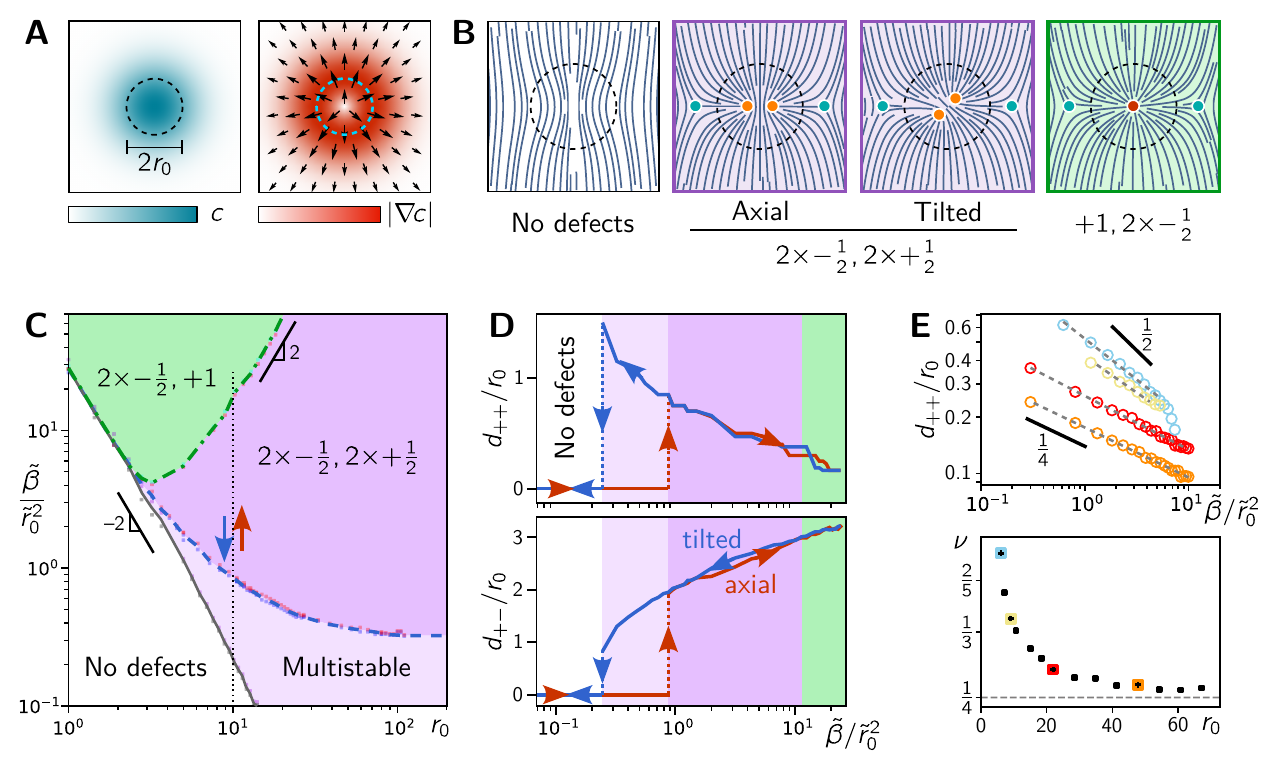}  
    \caption{
    Defect unbinding in planar geometry.
    \textbf{A} Gaussian concentration profile $c(r)$ and its gradient $\nabla c(r)$. The gradient magnitude is maximal at $r = r_0$.
    \textbf{B} Sketches illustrating the steady state configurations. 
    \textbf{C} Phase diagram of steady state configurations in the $(\tilde{r}_0, \tilde{\beta}/\tilde{r}_0^2)$ parameter plane. In the multistable region, the defect-free configuration and the unbound defect pairs coexist.
    \textbf{D} Defect separations as a function of $\tilde{\beta}$ from an adiabatic parameter sweep at $\tilde{r}_0 = 10$ (dotted black line in C; see Movie~2).
    \textbf{E} Power law scaling of the $+\frac12$ defect separation as a function of $\tilde{\beta}$ for different morphogen ranges $\tilde{r}_0$. For large $\tilde{r}_0$, the exponent $\nu$ approaches the theoretically predicted value $1/4$ (gray dashed line).
    }
    \label{fig:planar}
\end{figure*}

As the system is invariant under $\beta \to -\beta$, $\mathbf{Q} \to -\mathbf{Q}$ (corresponding by rotating the director by $\pi/2$), we will set $\beta > 0$ for the remainder of this paper.
Moreover, we will focus on radially symmetric morphogen concentrations with a Gaussian profile $c(r) = c_0 \exp[-r^2/(2r_0^2)]$.
The precise form of the profile is not important as long as it decays monotonically with a characteristic length scale $r_0$. For instance, the results do not change qualitatively if one uses an exponential gradient that result from a source-degradation-diffusion process \cite{Wartlick.etal2009}. (The case of a profile without a characteristic length scale as obtained from a purely diffusive process is discussed in the Supplementary Material.)

We can fix a length scale by expressing lengths in units of the nematic coherence length (defect core size) $\xi = \sqrt{K}$, and absorb a concentration scale factor $c_0$ and the elastic constant $K$ into a dimensionless alignment strength $\tilde{\beta} = \beta c_0^2 / K$.
Thus, the system has two dimensionless control parameters, alignment strength $\tilde{\beta}$ and the morphogen range $\tilde{r}_0$.

\textbf{\sffamily Defect unbinding, merging, and recombination.}
The director field configuration that conforms with the radial concentration gradient $\nabla c$ and therefore minimizes the alignment energy is an aster-like configuration with a $+1$ defect at the origin. 
However, such configuration incurs an elastic energy, since neighboring directors are not parallel to one another.
This implies that the elastic energy and the gradient-alignment compete with each other.
This competition is at the core of the phenomena we study in the following.
For vanishing gradient-alignment strength ($\tilde{\beta} = 0$) the steady state is a uniform, defect-free director field.
For a sufficiently large $\tilde{\beta}$, we expect that the alignment energy dominates resulting in the aster-like state with a $+1$ defect at the origin.
Defect charge conservation requires, however, that the $+1$ defect charge is balanced by negatively charged defects with net charge $-1$. We therefore expect that a defect-unbinding transition will take place with increasing alignment strength $\tilde{\beta}$. We expect that this unbinding takes place near $r = r_0$, where the gradient magnitude $|\nabla c|^2$ is maximal (dashed circles in Fig.~\ref{fig:planar}A).

To map out the behavior in the $(\tilde{r}_0, \tilde{\beta})$-parameter plane, we solve the relaxational dynamics of Eq.~\eqref{eq:free-energy-plane} (see Methods) using a Finite Element Method (FEniCS) in a square domain with free (Neumann) boundary conditions and a side length $L \gg r_0$ chosen large enough to avoid finite size effects.
For each morphogen range $\tilde{r}_0$, we initialize the system at $\tilde{\beta} = 0$ in a uniform state and then adiabatically increase the coupling strength $\tilde{\beta}$ (red arrows in Fig.~\ref{fig:planar}C,D). 
For small $\tilde{\beta}$, the director field can minimize its free energy by bending slightly toward the direction of the morphogen gradient around $\tilde{r} \approx \tilde{r}_0$, while remaining defect free (see Fig.~\ref{fig:planar}B and Movie~2).
Further increasing the coupling strength induces more bending which increases the elastic energy.
Above a critical coupling strength $\tilde{\beta}^*_\mathrm{unbind}$ (dashed blue line), this trade-off causes unbinding of two $\pm\frac{1}{2}$ defect pairs, opposite to each other, along an axis perpendicular to the initial director field.
For large $\tilde{r}_0$ the unbinding transition asymptotically scales as $\tilde{\beta}^*_\mathrm{unbind} \sim \tilde{r}_0^2$ (see Fig.~\ref{fig:planar}C).
This scaling can be rationalized by equating the energy required for defect unbinding, $E_\mathrm{unbind}\sim K$, to the alignment energy from the core region (with area $\mathcal{O}(\xi^2)$) which is largest at at $r = r_0$ and scales as $E_\mathrm{align} \sim \beta \xi^2/r_0^2$. 

To further quantify the unbinding, we measure the separation, $d_{+-}$, between each $\pm\frac12$ pair and the separation, $d_{++}$, between the two $+\frac{1}{2}$ defects (see Fig.~\ref{fig:planar}D).
Notably, after unbinding $d_{+-}$ immediately jumps to a finite value ($\sim 1.8 r_0$), indicating that the unbinding transition is discontinuous.
As $\tilde{\beta}$ is increased further, the distance $d_{+-}$ continues to increase while $d_{++}$ decreases until $d_{++} < 2 \xi$ and the $+\frac{1}{2}$ defect cores overlap, corresponding to their merging into a $+1$ defect at the origin. Unbinding of neutral pairs occurs where $|\nabla c|$ is largest, i.e.\ at $r \approx r_0$ such that, upon unbinding, the $+\frac{1}{2}$ defects are initially separated by $2 r_0$. Therefore, if $r_0 < 2 \xi$ unbinding immediately results in the $+1, 2{\times}{-}\frac{1}{2}$ configuration. For $r_0 \gg 2 \xi$, a critical alignment strength $\tilde{\beta}^*_\mathrm{merge}$ is required for the merging of the two $+\frac12$ into a $+1$ defect. This value increases with $\tilde{r}_0$ (dot-dashed green line) and asymptotically scales as $\tilde{\beta}^*_\mathrm{merge} \sim \tilde{r}_0^4$. This scaling 
arises from the asymptotic behavior of the $+\frac{1}{2}$ defect separation $d_{++}/r_0 \sim \tilde{\beta}^{-1/4}$ (see below).

Reversing the direction of the adiabatic sweep (decreasing $\tilde{\beta}$, blue arrow), the $+1$ defect splits into two $+\frac{1}{2}$ defects. These defects are tilted relative to the $-\frac{1}{2}$ defects (see Fig.~\ref{fig:planar}B). This slightly increases $d_{+-}$ compared to the forward sweep. 
Upon further decreasing $\tilde{\beta}$, the $\pm\frac12$ defect pairs show hysteresis: they persist beyond the unbinding transition
and recombine at a value of $\tilde{\beta}<\tilde{\beta}^*_\mathrm{unbind.}$. 
Hence, unbinding and recombination define a hysteresis loop encircling a bistable region, where the defect-free state and the defect-pair state coexist. Which of these two states is reached in the steady state depends on the initial condition.
Notably, the recombination threshold (gray solid line in Fig.~\ref{fig:planar}B) is given by a constant value of $\tilde{\beta} \approx 12$. The locus of the saddle-node bifurcation where the $2 \times \pm\frac12$ are annihilated can be estimated by a perturbation analysis which yields $\tilde{\beta}^*_\mathrm{recomb.} \approx 5.4$ (see SI Appendix).

In \textit{Hydra}, tentacle formation and budding require the de-novo formation of a $+1$ defect flanked by two $-\frac12$ defects in the actin orientation. These defects will sit at the tip and the base of the future tentacle/bud, respectively \cite{Philipp.etal2009,Aufschnaiter.etal2017}. Spot-like morphogen concentrations determine the sites of tentacle and bud formation \cite{Philipp.etal2009,Lengfeld.etal2009, Meinhardt1993}, suggesting that gradient alignment-induced defect unbinding might drive the actin reorganization during these morphogenetic processes (cf.\ Fig.~\ref{fig:overview}F). 

\textbf{\sffamily Asymptotic scaling of $+\frac12$ defect separation.}
After defect unbinding, numerical simulations show that gradient alignment drives the two $+\frac{1}{2}$ defects towards the concentration maximum where the director is aligned with the radial gradient. The steady-state defect separation is set by the balance of the gradient-alignment force and the repulsive force of the $+\frac{1}{2}$ defects.

To estimate the defect separation, we consider the regime $d_{++}/2 \ll r_0$ which allows us to use a perturbative approach.
When $\beta = 0$ the configuration of the director field $\mathbf{n} = (\cos \psi, \sin \psi)$ is governed by the elastic energy $E_\mathrm{el}=\frac{K}{2}\int\! \mathrm{d}x^2 |\nabla \mathbf{n}|^2 = \frac{K}{2}\int\! \mathrm{d}x^2 |\nabla \psi|^2$. The director angle, $\psi$, that minimizes this energy for two $+\frac{1}{2}$ defects at a distance $d$ is given by
\begin{equation} \label{eq:director-field}
    \psi(x,y) = \frac{1}{2} \arctan (x - d/2, y) + \frac{1}{2} \arctan (x + d/2, y)\;,
\end{equation}
and has an elastic energy $E_\mathrm{el} = -\frac{\pi K}{2} \, \log(d/\xi)$ which is analogous to the electrostatic energy of a pair of electric charges.
To estimate the energy of  alignment with the gradient, $E_\mathrm{a}$, we consider the regions near the defects and far from the defects separately:
$E_\mathrm{a} = E_\mathrm{a}^< + E_\mathrm{a}^> = \int_0^\lambda \mathrm{d}r \int \! \mathrm{d}\phi f_\mathrm{a} + \int_\lambda^\infty \mathrm{d} r \int \! \mathrm{d} \phi f_\mathrm{a}$, where $\lambda$ is an intermediate scale $d/2 \ll \lambda \ll r_0$.
In the following we perform a scaling analysis. The full calculation is presented in the SI Appendix.
In the outer region the deviation of the director orientation from the radial gradient $\delta \psi = \psi - \phi \sim (d/r)^2$ is small
and hence $f_\mathrm{a} = \beta |\nabla c|^2 \cos(\delta \psi)^2 \approx \beta |\nabla c|^2 (d/r)^4 + \text{const}$.
Since $|\nabla c|^2$ reaches its maximum value $c_0^2/r_0^2$ at $r = r_0$, we can estimate $E_\mathrm{a}^> \sim \beta c_0^2 (d/r_0)^4$.
On the other hand, near the defects ($r \approx d$)  $\delta \psi = \mathcal{O}(1)$ and 
${|\nabla c|^2 \approx c_0^2 r^2/r_0^4}$. Therefore, the director field is only weakly perturbed by the alignment interaction, which justifies estimating $E_\mathrm{a}$ and $E_\mathrm{el}$ using the unperturbed director field Eq.~\eqref{eq:director-field}.
The alignment energy in the inner region can then be estimated as $E_\mathrm{a}^< \sim \beta c_0^2 \int_0^d \mathrm{d}r r \, r^2/r_0^4 \sim \beta c_0^2 (d/r_0)^4$ and exhibits the same scaling as the outer contribution, $E_\mathrm{a}^>$.
Hence, $F_\mathrm{a} = -\partial_d E_\mathrm{a} \sim -\beta c_0^2 d^3/r_0^4$.
The defect separation in steady state is such that the alignment-mediated force balances the elastic repulsion $F_\mathrm{el} = -\partial_d E_\mathrm{el} \sim K/d$. Therefore, in steady state, we expect $d_{++}/r_0 \sim (\beta c_0^2/K)^{-1/4} = \tilde{\beta}^{-1/4}$.

We tested this prediction by fitting a power law $d_{++}/r_0 \sim \tilde{\beta}^{-\nu}$ to data from numerical simulations
[see Fig.~\ref{fig:planar}(e)]. For large $r_0$ we find that $\nu$ approaches $1/4$ as predicted from perturbation theory. For small morphogen range $r_0$, the power law exponent $\nu$ increases and eventually the scaling breaks down once the defect cores start overlapping, i.e.\ $d \approx \xi$.

\textbf{\sffamily Curved surfaces.}
Biological tissues, such as the ectoderm of regenerating \textit{Hydra}, often form closed surfaces that are have the topology of a sphere. This has two important consequences for nematic textures on the surface. First, the topology of a sphere necessitates a net defect charge of $+2$.  
Second, curvature modifies the nematic free energy and topological defects are attracted to regions with like-signed Gaussian curvature \cite{Bowick.Giomi2009}. Motivated by the evolution of \textit{Hydra}'s  body shape during regeneration, we study the interplay between changes in texture induced by geometry and topology and those driven by a morphogen gradient on curved surfaces.
We generalize the planar elastic free energy to curved surfaces in a minimal way by replacing the partial derivatives with the covariant derivatives. Including coupling to extrinsic curvature \cite{Napoli.Vergori2012,Biscari.Terentjev2006} and going beyond the one elastic constant approximation \cite{Shin.etal2008} are beyond the scope of this work, but will be interesting direction for future research.

\begin{figure*}
    \centering
    \includegraphics{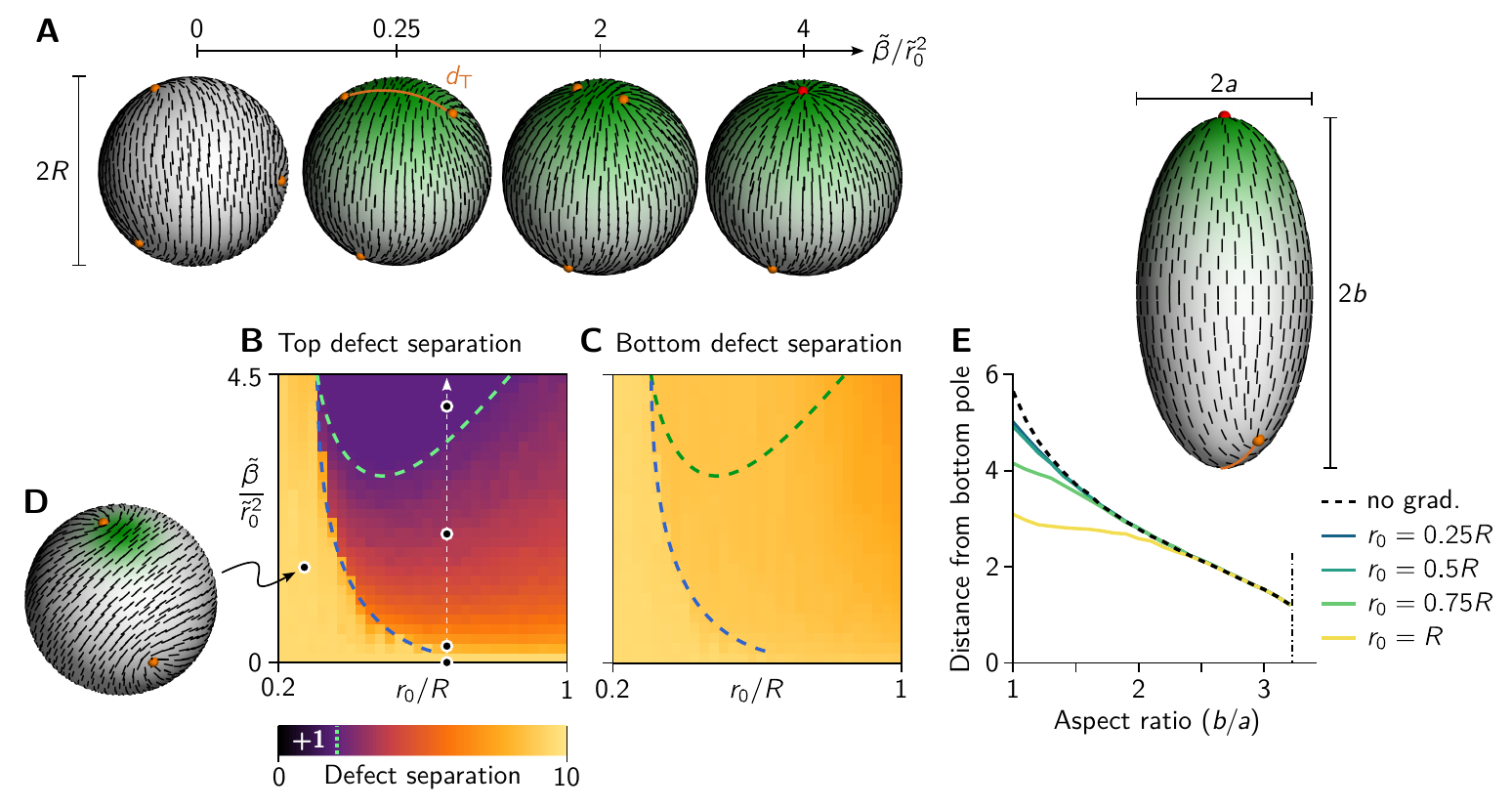}
    \caption{
    Gradient alignment drives topological defect positioning and merger on spherical and ellipsoidal shells.
    \textbf{A} Defect configurations on a sphere with radius $R = 6 \xi$ for increasing gradient alignment strength $\tilde{\beta}$.
    \textbf{B,C} Separation of top (B) and bottom (C) defect pairs, $d_\mathrm{T}$, $d_\mathrm{B}$, as a function of $\tilde{\beta}/\tilde{r}_0^2$ and $\tilde{r}_0$. For defect separations below the defect core size, the pair of $+\frac12$ defects is classified as a $+1$ defect (region above green dashed line). For small gradient ranges (region left of the blue dashed line), only one defect is pinned by the morphogen gradient while the other defects remain in an approximately tetrahedral configuration (\textbf{D}). As a consequence, the defect distance remains approximately constant despite the increasing alignment strength. The bottom defect separation is only weakly affected by gradient alignment.
    \textbf{E} Quantification of the bottom defects to the bottom pole on ellipsoids with varying aspect ratios $\alpha = b/a$. Both gradient alignment and the Gaussian curvature gradient drives the bottom two defects closer to the pole. For aspect ratios $\alpha \gtrsim 2$, curvature effects dominate. For aspect ratios $\alpha > 3.3$ (vertical dot-dashed line), the defects merge to a $+1$ defect at the pole. (The alignment strength was set to ensure a $+1$ defect at the top: $\tilde{\beta} = 10 \, \tilde{r}_0^2$.)
    }
    \label{fig:sphere-and-ellipsoid}
\end{figure*}

Let us start with a spherical surface, where, in the absence of gradient alignment, a nematic liquid crystal has four $+\frac12$ defects arranged in a tetrahedral configuration that maximizes their separation (leftmost panel in Fig.~\ref{fig:sphere-and-ellipsoid}A) \cite{Shin.etal2008, Napoli.Vergori2021}.

The spherical symmetry allows this configuration to be freely rotated.
Any small gradient alignment breaks the symmetry, causing the tetrahedral defect arrangement to rotate such that one of the $+\frac12$ defects is pinned at the location of maximum $|\nabla c|$, with its tail pointing down gradient (see Fig.~\ref{fig:sphere-and-ellipsoid}D).
For sufficiently small values of the range of the morphogen gradient ($r_0 \gtrsim R/3$), only one defect is attracted while the defects remain in a tetrahedral configuration relative to each other even as the alignment strength is increased (see Fig.~\ref{fig:sphere-and-ellipsoid}B)

For larger $r_0$, in contrast, the gradient alignment attracts two defects simultaneously, causing them to converge toward the concentration maximum with increasing $\tilde{\beta}$. Eventually they merging into a $+1$ defect as their separation drops below the defect core size. 

This suggests that the migration and eventual merger of two $+\frac12$ defects at the future head of the animal is driven by alignment to gradients of the head organizer morphogen.
Alternatively, for sufficiently strong gradient alignment, a $+1$ defect can directly form from an initially disordered state.
Importantly, the range of the morphogen gradient needs to be sufficiently large ($r_0 \gtrsim R/3$) to capture two $+\frac12$ defects and induce the formation of a $+1$ defect for strong alignment.

The remaining two $+\frac12$ defects, in contrast, move only slightly toward the opposite pole and remain well separated for all $\tilde{\beta}$ (see Fig.~\ref{fig:sphere-and-ellipsoid}D). 
This is in agreement with experimental observations showing that the merger of the $+\frac12$ defects near the foot takes place at a much later stage where the body shape has attained significant non-uniform curvature \cite{Maroudas-Sacks.etal2021}.
Previous analytical work \cite{Kamien2002,Vitelli.Nelson2004,Bowick.Giomi2009,Kralj.etal2011} has shown that topological defects are attracted to regions of like-signed Gaussian curvature. We therefore hypothesize that, while the merger of defects at the head is driven by morphogen gradients,  the migration and merger of the $+\frac12$ defects near the foot is a consequence of geometrical changes and increased curvature. 
To test this, we performed simulations on ellipsoids of varying aspect ratio $\alpha = b/a$, while keeping the surface area constant.
For $\alpha \lesssim 2$, the  morphogen gradient centered at the future location of the head repels the bottom two defects, driving them slightly closer towards the pole (Fig.~\ref{fig:sphere-and-ellipsoid}E).
For larger aspect ratios ($\alpha \gtrsim 2$), the curvature effect near the pole of the ellipsoid dominates and the defect separation becomes independent of $r_0$, confirming our hypothesis that the merging of the $+\frac12$ defects at the foot of \textit{Hydra} is driven by the increased curvature there, independently of the head organizer gradient. Of course this does not exclude the possibility that an additional  ``foot organizer'' morphogen gradient may drive the defect merger at the foot.

\begin{figure}
    \centering
    \includegraphics{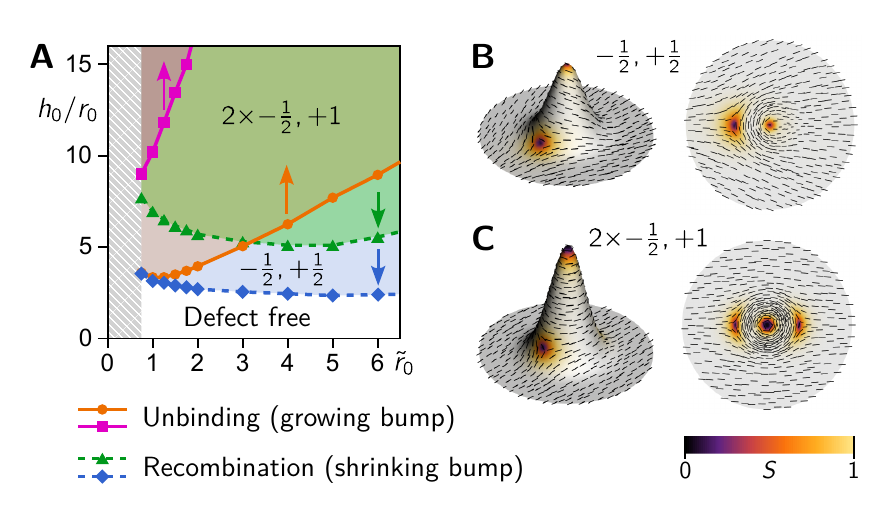}
    \caption{Curvature driven defect unbinding.
    \textbf{A} Phase diagram of defect unbinding and recombination transitions in the $\tilde{r}_0$-$h_0/r_0$ parameter plane. Solid lines show the bump heights where $\pm\frac12$ defect pairs unbind on an adiabatically growing bump (see Movie~3). Starting from the defect free configuration, the first unbinding transition (orange line, $\bullet$) gives rise to a $+\frac12$ defect at the tip and a $-\frac12$ defect at the base (\textbf{B}). The second unbinding transition (magenta line, $\scriptstyle\blacksquare$) results in a parity symmetric configuration with a $+1$ defect at the tip flanked by $-\frac12$ defects at the base (\textbf{C}). The corresponding recombination transitions (blue and green dashed lines) on an adiabatically shrinking bump take place at lower bump heights, indicating hysteresis (see Movie~4).
    }
    \label{fig:gaussian-bump}
\end{figure}

The above results suggest that gradient alignment and non-uniform curvature have similar effects on existing defects. With this in mind, we return to the question of unbinding of new defects and ask what the role of curvature might be during budding and tentacle formation. Indeed, localized Gaussian curvature has been shown to drive unbinding of topological defects in the context of crystal dislocations and superfluid vortices \cite{Vitelli.Nelson2004,Vitelli.etal2006,Turner.etal2010,Jimenez.etal2016}.
In analogy with the Gaussian concentration profile, we performed simulations on surfaces with a Gaussian-shaped height profile $h(r) = h_0 \exp[-r^2/(2 r_0^2)]$; see Fig.~\ref{fig:gaussian-bump}. Starting from a uniform configuration on a planar surface, the bump height is adiabatically increased (see Methods for details). Above a critical bump height, a $\pm\frac12$ defect pair unbinds, with the $+\frac12$ defect localized at the tip of the bump, where the Gaussian curvature is positive, and the $-\frac12$ defect near the base, where the Gaussian curvature is negative (Fig.~\ref{fig:gaussian-bump}B and Movie~3).
Only for larger bump heights, a second unbinding occurs forming a $+1$ defect at tip and an additional $-\frac12$ at the base (Fig.~\ref{fig:gaussian-bump}C). This is in marked contrast to gradient alignment which induces unbinding of two $\pm\frac12$ pairs (cf.\ Fig.~\ref{fig:planar}B,C) without an intermediate regime of a single $\pm\frac12$ pair.
The reason for this difference is that Gaussian curvature drives defect unbinding as a consequence of angular deficiency (i.e.\ angles don't add up to $360^\circ$ when going around a circle) without imposing a preferred director orientation. The angular deficiency can be accommodated by first unbinding a single $\pm\frac12$ defect pair. 
By contrast, alignment to a radial gradient is accommodated by a radially oriented director field which requires either a $+1$ defect or a pair of $+\frac12$ defects.

Unbinding of a single $\pm\frac12$ defect pair has not been observed during budding and tentacle formation in \textit{Hydra}, suggesting that curvature may play only a minor role in the initial reorganization of myonemes during these morphogenetic processes. During later stages, curvature can help to stabilize the $+1$ defects at the tip of a bud or tentacle (cf.\ Fig.~\ref{fig:overview}E).
In future work, further analytic insight into the curvature-induced unbinding transition could be gained using the methods developed in Ref.~\cite{Vitelli.Nelson2004}.

\section{\sffamily Discussion}
Our minimal model shows how coupling of nematic orientational order to a morphogen gradient and to curvature can together recapitulate the nematic organization of actin fibers in \textit{Hydra} during regeneration (Fig.~\ref{fig:overview}B--E) and budding (Fig.~\ref{fig:overview}F). Specifically, we show that these mechanisms can overcome the elastic energy that prefers parallel alignment of fibers, driving both the merging of two $+\frac12$ defects into a $+1$ and defect unbinding from a uniform background.
Our results suggest that alignment to morphogen gradients, rather than curvature effects, takes the leading role in these processes.
Once the body morphology has been established, curvature can then stabilize topological defects.

Here, we have examined a minimal model which lays the foundation for future extensions.
Going forward, it will be interesting to go beyond the one Frank elastic constant approximation to account for different responses to bend, splay, twist, and biaxial splay deformations \cite{Selinger2018,Shin.etal2008}. 
Other natural extensions include coupling to extrinsic curvature \cite{Napoli.Vergori2012,Biscari.Terentjev2006}, endowing the morphogen field(s) with dynamics and feedback from texture deformations \cite{Meinhardt1993,Maryshev.etal2020,Mercker.etal2021,Norton.Grover2022}.
It will also be interesting to study gradient alignment on deforming elastic shells, where the nematic order determines the local stresses that drive deformations \cite{Metselaar.etal2019,Mietke.etal2019,Heer.etal2017,Duffy.Biggins2020,Vafa.Mahadevan2022,Khoromskaia.Salbreux2021,Hoffmann.etal2022b}.
Since defects act as organizational centers for anisotropy fields, they naturally emerge as points with special mechanical properties, such as a concentration of active stresses \cite{Copenhagen.etal2021,Guillamat.etal2022}.

From a broader perspective our work addresses the question of how the local gradient of scalar morphogen concentration fields can (re)organize spatial anisotropy encoded by vectorial and tensorial fields. We have shown how this reorganization is governed by the competition between alignment to the gradient and alignment to neighbors (elastic energy) while it is topologically constrained by defect-charge conservation.
The importance of orientational information encoded in anisotropy fields is evident in the mechanical programs underlying morphogenesis \cite{Streichan.etal2018,Lefebvre.etal2022,Heisler.etal2010,Stokkermans.etal2022,Mitchell.etal2022} and in cell polarity \cite{Eaton.Julicher2011,Devenport2014}.
We therefore expect that our findings will help understand how morphogens, tissue curvature and anisotropic fields are coupled in a broad range of systems.

\section{\sffamily Materials and Methods}

\textbf{\sffamily Relaxational dynamics.}
The relaxational dynamics of the Q-tensor field as it approaches a (local) energy minimum is given by
\begin{equation}
\gamma \partial_t\mathbf{Q} = -\left(\frac{\delta E}{\delta\mathbf{Q}}\right)^{\!\mathrm{ST}}
\label{eq:Qdot}
\end{equation}
$\gamma$ the damping coefficient and the superscript ST denotes the symmetric and traceless part of a tensor. Explicitly evaluating the functional derivative gives
\begin{multline}
    \gamma \partial_t Q_{ij} = -(2 \Tr\mathbf{Q}^2-1)Q_{ij} + K \, \nabla^2Q_{ij} \\
    - \beta \left[(\nabla_i c)(\nabla_j c)-\frac12\delta_{ij}|\nabla c|^2\right].
\end{multline}
The damping coefficient $\gamma$ defines the timescale $\tau = \gamma^{-1}$ of the relaxational dynamics. 

In the planar domain, the symmetric, traceless Q-tensor has only two degrees of freedom, i.e.\ it can be written as
\begin{equation}
    \mathbf{Q} = \begin{pmatrix} q_1 & q_2 \\ q_2 & -q_1 \end{pmatrix}.
\end{equation}
The resulting PDEs for $q_1, q_2$ were simulated using FEniCS.

\textbf{\sffamily Surface finite element implementation of Q-tensor model.}
We formulate the covariant Q-tensor model on a curved surface as a surface finite element problem following the approach introduced in Ref.~\cite{Nestler.etal2019}. The central idea is to represent tensor fields in the surface's tangent bundle by tensor fields in the embedding Euclidian space and penalize out-of-surface components during the time evolution.

Let $\mathcal{S}$ be a surface (two-dimensional manifold) embedded in three-dimensional Euclidian space $\mathbb{R}^3$. Locally, the surface is described in terms of a parametrization $\mathbf{X}(\xi_1, \xi_2) \in \mathcal{S}$. In the following, we will denote vectors in the embedding space, $\mathbb{R}^3$, in bold font. Vectors and tensors in the surface will be denoted by their components in the local parametrization. 

The local basis vectors $\evec_i := \partial_i \mathbf{X} = \partial_{\xi_i} \mathbf{X}$, $i = 1,2$, span the local tangent space. We can now find the metric tensor $g_{ij} = \evec_i \cdot \evec_j$, where $\cdot$ denotes the standard inner product in $\mathbb{R}^3$. The dual basis vectors $\evec_i$ are defined via the orthonormality conditions $\evec^i \cdot \evec_j = \delta^i_j$.
The surface normal vector field is given by
\begin{equation}
    \bm{\nu} = \frac{\evec_1 \times \evec_2}{|\evec_1 \times \evec_2|} \; .
\end{equation}
Using the normal vector field, we can define the second fundamental form (curvature tensor) $b_{ij} = \bm{\nu} \cdot \partial_i \evec_j$.

Vector fields in the tangent bundle of $\mathcal{S}$ can be represented by vector fields in the embedding space via $\mathbf{u} = u^i \evec_i$ (and analogously for tensor fields).
This allows one to implement the numerical simulation without explicitly parametrizing the surface. 
However, due to numerical inaccuracies, the out-of-surface component of the representing vector will never be exactly zero. To account for this, we write the approximate representing vector as $\tilde{\mathbf{u}} = u^i \evec_i + \tilde{u}_n \bm \nu$.
To keep $\tilde{u}_n$ small, one introduces a penalty force $\mathbf{p}(\tilde{\mathbf{u}}) \propto \Pi[\tilde{\mathbf{u}}] - \tilde{\mathbf{u}} = (\tilde{\mathbf{u}} \cdot \bm{\nu}) \bm{\nu}$ in the dynamics. For a detailed discussion see Ref.~\cite{Nestler.etal2019}.

Given a vector $\tilde{\mathbf{u}}$ in the embedding space, the corresponding surface vector can be found by projection into the local tangent space with the projector 
\begin{equation}
    \Pi = \mathbb{I}_3 - \bm{\nu} \otimes \bm{\nu}.
\end{equation}
Going from the planar geometry to a curved surface, the gradient operator $\nabla$ needs to be replaced by the covariant derivative. For scalar fields, this is simply the partial derivative in the surface coordinates $\nabla \phi = \evec^i \partial_i \phi$. However, for vector and tensor fields, extra terms appear because of parallel transport:
\begin{equation}
    \nabla_{\!i} u^j = (\partial_i \mathbf{u}) \cdot \evec^j = \partial_i u^j + \Gamma_{ik}^j u^k,
\end{equation}
with the Christoffel symbols $\Gamma_{ij}^k = \Gamma_{ji}^k = (\partial_i \evec_j) \cdot \evec^k$. We use Einstein summation convention for repeated indices.
While finite element methods such as FEniCS supply the (componentwise, non-covariant) surface gradient operator $\tilde{\nabla}_{\!\mathcal{S}} := \evec^i \partial_i$, they do not supply covariant derivatives. To relate $\tilde{\nabla}_{\!\mathcal{S}}$ to $\nabla$, we explicitly calculate 
\begin{align}
    \tilde{\nabla}_{\!\mathcal{S}} \tilde{\mathbf{u}}
    &= \evec^i \otimes \partial_i (u^j \evec_j + \tilde{u}_n \bm{\nu}) \notag \\
    &= \left(\nabla_{\!i} u^j - \tilde{u}_n b_i^j \right) \evec^i \otimes \evec_j + \left(b_{ik} u^k + \partial_i \tilde{v}_n\right)\evec^i \otimes \bm{\nu}
\end{align}
where we used the relations $\partial_i \evec_j = \Gamma_{ij}^k \evec_k + b_{ij} \bm{\nu}$ and $\partial_i \bm{\nu} = - b_i^j \evec_j$.
We can now project to the tangent space and reorganize to find an expression for the covariant derivative in terms of the embedded vector
\begin{equation}
    (\nabla_{\!i} u^j) \, \evec^i \otimes \evec_j = \Pi[\tilde{\nabla}_{\!\mathcal{S}} \tilde{\mathbf{u}}] + (\bm{\nu} \cdot \tilde{\mathbf{u}}) \, \mathbf{B}
\end{equation}
where we introduced the curvature matrix $\mathbf{B} = b_i^j \, \evec^i \otimes \evec_j$. Enforcing tangentiality of $\tilde{\mathbf{u}}$ with a sufficiently strong normal penalty allows one to neglect the second term, which simplifies the numerical implementation. 
(Note that numerical errors due to spatial and temporal discretization will always lead to a small, non-zero out-of-surface component of $\tilde{\mathbf{v}}$. Therefore, higher numerical accuracy can be achieved by including the explicit curvature-coupling term \cite{Nestler.etal2019}.) 

\newcommand{\Qrep}{\tilde{\mathbf{Q}}}
\newcommand{\Qtest}{\Qrep_\mathrm{test}}
The above derivation generalizes to tensor fields $\mathbf{Q} = Q^{ij} \evec_i \otimes \evec_j$ and their approximate representation $\Qrep$, s.t.\ $\Pi[\Qrep] = \mathbf{Q}$.
The surface nematic tensor must be a traceless, symmetric tensor. In the surface, the trace is given by
\begin{align}
    0 \overset{!}{=} \Tr_\mathcal{S} \mathbf{Q} &= g^{ij} Q_{ij} = \evec_i \cdot \mathbf{Q} \cdot \evec_j \notag \\
    &= \evec_i \cdot \Pi[\Qrep] \cdot \evec_j = \Tr \Qrep - \bm{\nu} \cdot \Qrep \cdot \bm{\nu}
\end{align}
To ensure that $\Tr \Qrep = 0$ entails $\Tr_\mathcal{S} \mathbf{Q} = 0$, we define a new projector that projects in to the space of traceless surface tensors
\begin{equation}
    \Pi_\mathrm{T}[\Qrep] := \Pi[\Qrep] + \frac{1}{2} (\bm{\nu} \cdot \Qrep \cdot \bm{\nu}) \, \mathbf{g},
\end{equation}
where $\mathbf{g} = g_{ij} \, \evec^i \otimes \evec^j$.
The penalty force corresponding to this projector is given by
\begin{multline}
    \mathbf{P}_\mathrm{T}(\Qrep) \propto \Pi_\mathrm{T}[\Qrep] - \Qrep
    = \bm{\nu} \otimes (\bm{\nu} \cdot \Qrep) + (\Qrep \cdot \bm{\nu}) \otimes  \bm{\nu} \\
    - \frac{1}{2} (\bm{\nu} \cdot \Qrep \cdot \bm{\nu}) \, (\bm{\nu} \otimes \bm{\nu} + \mathbb{I}).
\end{multline}
The extra identity matrix in the last term makes sure that $\Tr \mathbf{P}_\mathrm{T} = 0$.

We can now write the dynamics in weak form, which can be straightforwardly implemented in a finite element solver:
\begin{multline} \label{eq:weak-form}
    \int_\mathcal{S} \mathrm{d}A \, (\partial_t \Qrep):\Qtest =
    \int_{\mathcal{S}} \mathrm{d}A \Big[
         (\Tr\Qrep^2-1) \Qrep : \Qtest \\
         + K \, \Pi[\tilde{\nabla}_{\!\mathcal{S}} \Qrep] : \Pi[\tilde{\nabla}_{\!\mathcal{S}} \Qtest] \\
         + \beta \, \mathbf{G} : \Qtest +
         \omega \, \mathbf{P}_\mathrm{T}(\Qrep) : \Qtest
     \Big],
\end{multline}
where $:$ denotes a full contraction of tensors. Since the projection operator $\Pi$ is idempotent, the elastic energy term can be simplified to $K \, \Pi[\tilde{\nabla}_{\!\mathcal{S}} \Qrep] \,{:}\, \tilde{\nabla}_{\!\mathcal{S}} \Qtest$. $\omega$ is the penalty strength which we set to 1000 \cite{Nestler.etal2019}.
$\Qtest$ is a test function from the space of symmetric, traceless tensor fields in $\mathbb{R}^3$ with appropriate continuity and differentiability constraints.
The tensor $\mathbf{G}$ is the traceless, symmetric tensor representation of the concentration gradient
\begin{equation}
    \mathbf{G} = \nabla c \otimes  \nabla c - \frac{\mathbb{I}_3}{3} |\nabla c|^2
\end{equation}

Since the traceless, symmetric Q-tensor has only 5 independent components, we can write it as
\begin{equation}
    \Qrep = \begin{bmatrix}
    q_1 & q_2 & q_3\\
    q_2 & q_4 & q_5\\
    q_3  & q_5 & -q_1-q_4
    \end{bmatrix}
\end{equation}
and analogously for $\Qtest$. The degrees of freedom in the numerical simulation are then $q_i$, $i = 1... 5$.

The time derivative on the left-hand side of Eq.~\eqref{eq:weak-form} is discretized using a simple forward (explicit) Euler scheme. 
The numerical simulations were implemented in Python 3.8 using FEniCS \cite{Alnaes.etal2015}.
Planar meshes were generated with FEniCS-mshr, and the curved surface meshes (sphere, ellipsoid, and Gaussian bump) were generated using the Gmsh-python API. The \textit{Hydra}-shaped surface was modeled in Blender and the exported mesh was cleaned up in MeshLab using the LS3Loop filter. The mesh sizes were chosen sufficiently small avoid discretization artefacts.

\textbf{\sffamily Ellipsoid geometry.}
In the parameter sweeps varying the aspect ratio of the ellipsoid, we keep the surface area fixed. This is motivated by the fact that there is no significant cell proliferation during regeneration of \textit{Hydra}, such that the surface of the epithelium remains approximately constant.

The surface area of a prolate ellipsoid with semi-minor axis $a$ and semi-major axis $b \geq a$ is given by
\begin{equation}
    A = 2\pi a^2\left(1 + \frac{b}{a \varepsilon}  \arcsin \varepsilon\right)
                   \quad\text{where } \varepsilon^2 = 1 - \frac{a^2}{b^2}
\end{equation}
Thus, a prolate ellipsoid with a given aspect ratio $\alpha = b/a$ and area $A = 4\pi R^2$ (equal to that of a reference sphere with radius $R$) has semi axes
\begin{equation}
    a = \frac{b}{\alpha} = \frac{\sqrt{2} R}{\sqrt{1+\frac{\alpha}{\varepsilon} \arcsin \varepsilon}} \qquad\text{where } \varepsilon^2 = 1-\frac{1}{\alpha^2}.
\end{equation}

\textbf{\sffamily Gaussian bump.}
We parametrize the radially symmetric Gaussian bump with height function $h(r) = h_0 \exp[-(x^2 + y^2)/(2 r_0^2)]$.
Adiabatic sweeps of the bump height were performed in COMSOL Multiphysics using the \texttt{Deformed Geometry} and \texttt{Auomatic Remeshing} features on a disk geometry with Neumann boundary conditions. For the growing bump sweeps, the director field was initialized uniformly aligned on a flat disk and the bump height was increased as $h_0(t)/r_0 = 0.5 \times 10^{-4} \, t$ (time, $t$, in units of the nematic relaxation time) up to $h_0 = 15 r_0$. For the shrinking bump, the initial height was set to $h_0(t = 0) = 10 \, r_0$, and the director field was initialized pointing along the $x$-axis. This initial configuration rapidly relaxes to a steady state with a $+1$ defect at the tip and two $-\frac12$ defects at the base. The bump height was adiabatically lowered [$h_0(t)/r_0 = 10 \, (1 - 0.5 \times 10^{4} \, t)$]. 

\textbf{\sffamily Code.} Available at \url{https://github.com/f-brauns/surface-nematic}

\textbf{\sffamily Acknowledgements.} The authors thank Eyal Karzbrun and Arthur Hernandez for inspiring discussions and Nikolas Claussen for critical feedback on the manuscript. FB was supported by Simons Foundation grant (\#216179) to the KITP. ZW and MCM were supported by the National Science Foundation award No. DMR-2041459.

\input{main.bbl}

\clearpage
\widetext

\newgeometry{total={5.5in,8.5in}}

\section{Supplementary material}

\setcounter{figure}{0}
\renewcommand\thefigure{S\arabic{figure}} 

\setcounter{equation}{0}
\renewcommand\theequation{S\arabic{equation}} 

\subsection{Scale-free concentration profile}

For purely diffusive concentration field with a point source in 2D, the steady state gradient has magnitude $\propto 1/r$, since the total diffusive flux through any circle around the origin must be conserved. The resulting concentration profile is logarithmic and hence has no characteristic length scale. (The singularity at the origin is of no concern since the source is never actually pointlike but has some finite size which regularizes the logarithm near the origin. In numerical simulations, we implement a Gaussian peak source which entails a gradient $(1 - e^{-r^2/(2 a^2)})/r$. The source size $a$ is chosen on the order of the nematic coherence length.) 
As the gradient magnitude is $1/r$, the alignment energy scales as $1/r^2$. Interestingly, this is the same scaling as the elastic energy density of a $+1$ defect at the origin. Therefore, for the logarithmic profile, alignment energy and elastic energy are on equal footing on the entire domain. This is in contrast to a profile with a finite length scale $r_0$, for which the elastic energy dominates for $r \gg r_0$. In the latter case, the $-\frac12$ defects generated due to charge conservation remain bounded.
By contrast, for a logarithmic concentration profile, the $-\frac12$ defects are expelled to the domain boundary where they are annihilated because we employed free (Neumann) boundary conditions (see Movie~5).

\subsection{Perturbation theory}

\paragraph{Defect separation.} To estimate the separation of $+\frac12$ defects in steady state and the defect recombination transition, we employ a perturbative ansatz using the steady-state director field in the absence of gradient alignment as a starting point. 
This director field $\mathbf{n}_0 = [\cos(\psi), \sin(\psi)]$ minimizes the elastic free energy
\begin{equation}
    E_\mathrm{el}[\mathbf{n}_0(x,y)] = \int \mathrm{d}x^2 \, \frac{K}{2} ||\nabla \mathbf{n}_0||^2 = \int \mathrm{d}x^2 \, \frac{K}{2} ||\nabla \psi||^2.
\end{equation}
Therefore, the minimizer obeys the harmonic equation 
\begin{equation} \label{eq:elastic-minimizer}
    0 = -\frac{\delta E_\mathrm{el}}{\delta \psi} = \nabla^2 \psi,
\end{equation}
with appropriate boundary conditions at the domain boundaries and around prescribed defects. 

On an infinite domain, with $+\frac12$ defects at $(\pm d/2, 0)$ the solution to Eq.~\eqref{eq:elastic-minimizer} is given by
\begin{equation} \label{eq:one-half-pair}
    \psi = \frac{1}{2} \arctan (x - d/2, y) + \frac{1}{2} \arctan (x + d/2, y),
\end{equation}
where the two-argument $\arctan(x,y)$ function gives the angle between the vector $(x,y)$ and the $x$-axis, taking into account the quadrant that the vector is in.
The elastic free energy of this configuration is found as $E_\mathrm{el}(d) = \frac{\pi K}{2} \log(d/\xi)$, where $\xi$ is the defect core radius at which the integration is cut of near the defects. 

The idea is now to use Eq.~\eqref{eq:one-half-pair} as an ansatz to estimate the alignment free energy $E_\mathrm{a}(d)$ a function of the defect separation $d$ and then find the steady state separation $d^*$ as the minimum of $E(d) = E_\mathrm{el}(d) + E_\mathrm{a}(d)$, i.e.\ solution to the force balance equation
\begin{equation}
    -\partial_d E(d^*) = F_\mathrm{el}(d^*) + F_\mathrm{a}(d^*) = 0.
\end{equation}
The elastic interaction is the well known ``Coulomb'' repulsion $F_\mathrm{el} = \pi K/(2d)$.
The alignment free energy to first order in $\beta$ can be calculated as
\begin{equation} \label{eq:alignment-energy}
    E_\mathrm{a}^{(1)} = \beta \int \mathrm{d}x^2 \, \left[1 - (\nabla c \cdot \mathbf{n}_0)^2 \right],
\end{equation}
where we have added an constant offset for convenience, such that $E_\mathrm{a}^{(1)} = 0$ when $\mathbf{n} \propto \nabla c$ everywhere.

\begin{figure*}[t]
    \centering
    \hspace*{-0.7in}
    \includegraphics{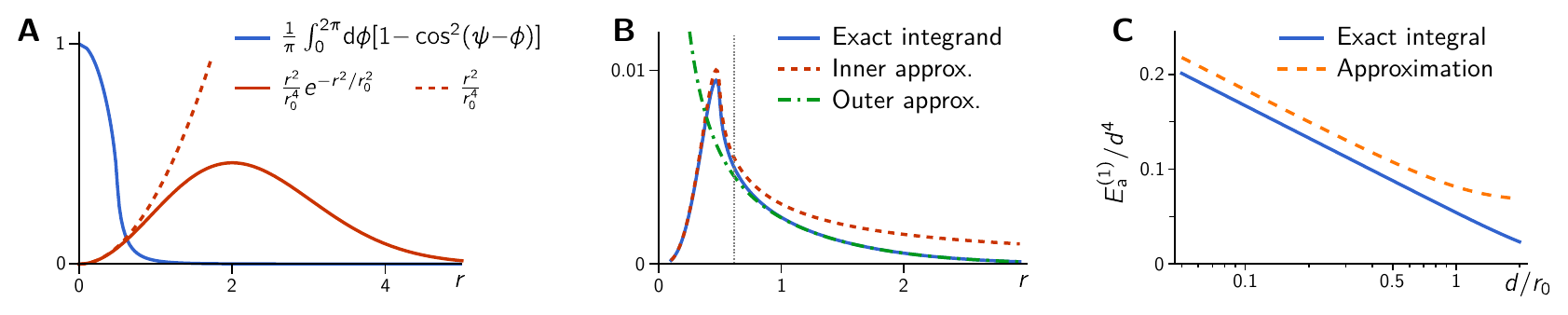}
    \caption{
    Approximation of the free energy integrals from the perturbation ansatz.
    \textbf{A} Two contributions to the radial integrand in the alignment energy: the azimuthally integrated ``angle deviation'' $I(r)$ [see Eq.~\eqref{eq:alignment-integral}] and the gradient magnitude $\partial_r c^2$. For $r \ll r_0$, the gradient magnitude is well approximated by $r^2/r_0^2$.
    \textbf{B} Separate approximations for radial integrand can be made for $r \ll d/2$ (red dashed line) and $r \gg d/2$ (green dash-dotted line). The numerically evaluated full integrand is shown for reference (solid blue line).
    \textbf{C} Numerically evaluated full integral Eq.~\eqref{eq:alignment-integral} vs approximation Eq.~\eqref{eq:alignment-energy-approx}, both scaled by $d^{-4}$ to emphasize the logarithmic deviation from the power law scaling.
    }
    \label{fig:integral-approx}
\end{figure*}

For a radial gradient, $\nabla c \propto (x,y)$, it is useful to change to polar coordinates $(r, \phi)$
\begin{align} \label{eq:alignment-integral}
    E_\mathrm{a}^{(1)} = \beta \int_0^\infty \mathrm{d}r \, r |\partial_r c|^2
    \underbrace{
    \int_0^{2\pi} \mathrm{d}\phi \,  \left[ 1 - \cos^2\bigl(\psi(r,\phi; d\bigr) - \phi)\right]
    }_{\displaystyle I(r)}.
\end{align}
For the Gaussian profile $c(r)$ used in the main text, the gradient magnitude is $|\partial_r c|^2 = c_0^2 \frac{r^2}{r_0^4} \exp(-r^2/r_0^2)$, which reaches it's maximum value of $1/r_0^2$ at $r = r_0$.
To make further progress, we approximate Eq.~\eqref{eq:alignment-energy} in the limit $r_0 \gg d/2$. 
Observe that for $r \approx d$, the gradient magnitude small and well approximated by $|\partial_r c|^2 \approx r^2/r_0^4$. On the other hand, for $r \approx r_0 \gg d$, where $|\partial_r c|^2$ is large, the deviation of the director from a radial orientation is $\delta \psi = \psi - \phi$ small. This suggests to split the radial integral into an inner and an outer integral at some intermediate scale $d/2 \ll \lambda \ll r_0$
\begin{equation}
    E_\mathrm{a}^{(1)} = E_\mathrm{a}^< + E_\mathrm{a}^>
\end{equation}
where we integrate $r \in [0, \lambda]$ in $E_\mathrm{a}^<$ and $r \in [\lambda, \infty]$ in $E_\mathrm{a}^<$. We can then use $|\partial_r c|^2 \approx c_0^2 r^2/r_0^4$ to approximate the integrand for the inner integral
\begin{align}
    E_\mathrm{a}^<
    &\approx \beta c_0^2 \int_0^\lambda \mathrm{d}r \, \frac{r^3}{r_0^4} \int_0^{2\pi} \mathrm{d}\phi \,  \left[ 1-\cos^2(\psi(r,\phi; d) - \phi) \right] \\
    &= \beta c_0^2 \frac{d^4}{r_0^4}
    \underbrace{
    \int_0^{\lambda/d} \mathrm{d}\rho \, \rho^3 \int_0^{2\pi} \mathrm{d}\phi \, \left[ 1 - \cos^2(\psi(\rho,\phi) - \phi) \right]
    }_{\displaystyle I(\lambda/d)}
\end{align}
where we used $\psi(r,\phi; d) = \psi(r/d,\phi)$ and made the substitution $r/d \to \rho$ in the second line.
If one now fixes the ratio $\lambda/d = \eta$ as a constant independent of $d$, the integral $I(\eta)$ yields a constant factor that can be evaluated numerically. For $\eta = 1$, we find $I \approx 0.22$.

For the outer integral, we expand Eq.~\eqref{eq:one-half-pair} to lowest order in $1/r$, giving $\psi \approx \phi + \frac{d^2}{8 r^2} \sin(2 \phi)$. After expanding $\cos^2(\delta \psi) = 1 - \delta \psi^2$ we find
\begin{align}
    E_\mathrm{a}^> &\approx \beta c_0^2 \int_{\eta d}^\infty \mathrm{d}r \, \frac{r^3}{r_0^4} e^{-r^2/r_0^2} \int_0^{2\pi} \mathrm{d}\phi \, \left(\frac{d^2}{8 r^2} \sin(2 \phi) \right)^2 \\
    &= \beta c_0^2 \frac{\pi}{64} \frac{d^4}{r_0^4} \int_{\eta d}^\infty \mathrm{d}r \, \frac{1}{r} e^{-r^2/r_0^2} \\
    &= \beta c_0^2 \, \frac{\pi}{128} \frac{d^4}{r_0^4} \, \Gamma \! \left(\!0, \frac{\eta^2 d^2}{2 r_0^2} \right),
\end{align}
where $\Gamma$ is the incomplete gamma function. Note that $\Gamma(0,x) \approx -\frac12 - \log x$ for $x \ll 1$. Therefore, the outer contribution to the free energy, $E_\mathrm{a}^>$, contributes a logarithmic correction for $d/2 \ll r_0$

Adding the inner and outer contributions we find 
\begin{equation} \label{eq:alignment-energy-approx}
    E_\mathrm{a}^{(1)} = \beta c_0^2 \, \frac{d^4}{r_0^4} \, \left[
    I(\eta) + 
    \frac{\pi}{128} \Gamma \! \left(\!0, \frac{\eta^2 d^2}{2 r_0^2} \right)
    \right] +
    \mathcal{O}\left(\frac{d^5}{r_0^5} \right),
\end{equation}
This approximation is compared the numerically evaluated integral Eq.~\eqref{eq:alignment-integral} in Fig.~\ref{fig:integral-approx}. The plot clearly shows that the $\Gamma$-function correction in Eq.~\eqref{eq:alignment-energy-approx} correctly captures the logarithmic deviation of $E_\mathrm{a}^{(1)}$ from a $d^4$ scaling.

\begin{figure}
    \centering
    \includegraphics{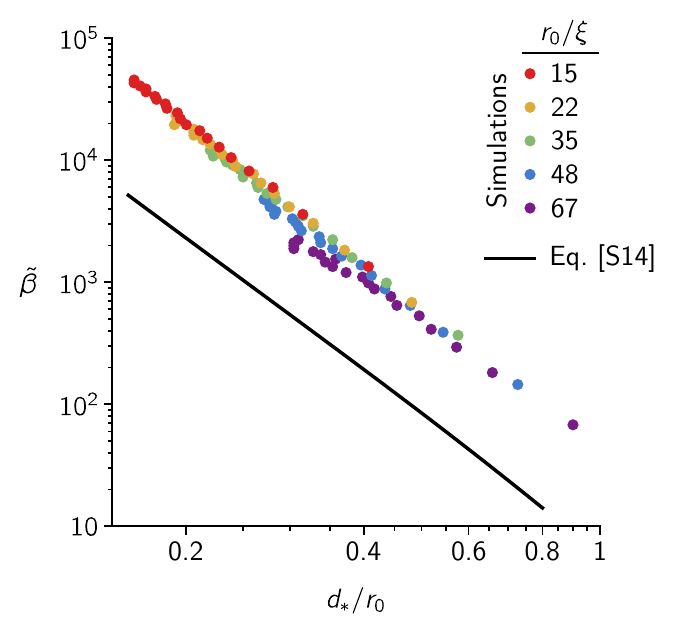}
    \caption{$+\frac12$ defect separation, numerical simulations and analytic estimate from perturbation theory.}
    \label{fig:defect-separation-numerics}
\end{figure}

Calculating $F_\mathrm{a}^{(1)} = -\partial_d E_\mathrm{a}^{(1)}$ and substituting into the force balance $F_\mathrm{el} + F_\mathrm{a} = 0$ one obtains an implicit equation for the defect separation $d_*$ in steady state
\begin{equation}
    \tilde{\beta}^{-1} = \frac{\pi d_*^4}{2 r_0^4} \, \left\{
    4 I(\eta) + 
    \frac{\pi}{32} \left[
        \Gamma \! \left(\!0, \frac{\eta^2 d_*^2}{2 r_0^2} \right) -
        \frac{1}{2} e^{-\eta^2 d_*^2/(2 r_0^2)} 
    \right]
    \right\}.
\end{equation}
The solution $d_*(\tilde{\beta})$ to this equation is obtained graphically by plotting $\tilde{\beta}(d_*)$ and inverting the graph.
Figure \ref{fig:defect-separation-numerics} shows the comparison to full numerical simulations. The perturbation theory captures the qualitative dependence of the the defect separation on $\tilde{\beta}$ but underestimates the defect separation by a factor $\sim 1.8$.

\paragraph{Defect recombination.}

To study the defect recombination transition we make an ansatz using the director field that minimizes the elastic free energy with prescribed defect positions. In other words, we ignore distortions of the director field by the gradient alignment. While this is not a well-controlled approximation, we can still hope to recover the qualitative nature of the defect-recombination transition.

\begin{figure}
    \centering
    \includegraphics{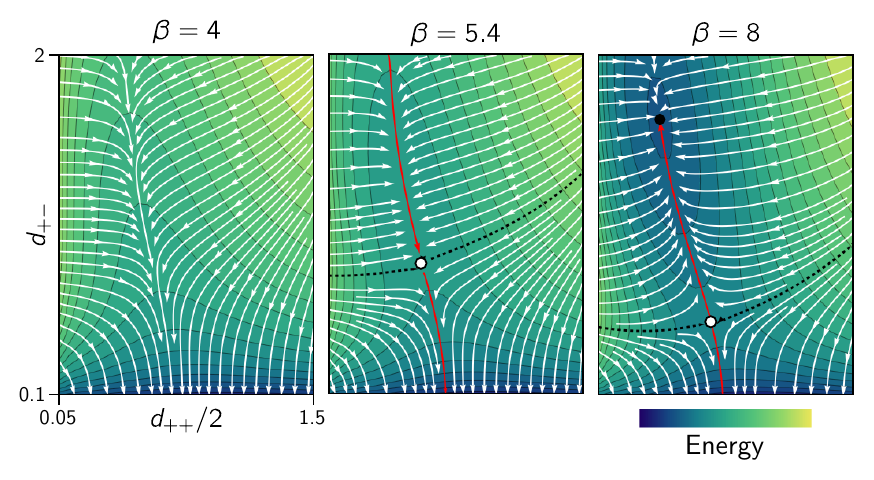}
    \caption{Free energy landscapes obtained from the perturbation ansatz with two pairs of $\pm\frac12$ defects as a function of the defect separations $d_{++}$ and $d_{+-}$. The limit $d_{+-} \to 0$ corresponds to the defect free state. White arrows show the gradient flow in the energy landscape. For $\tilde{\beta} < 5.4$, the flow goes toward $d_{+-} \to 0$, leading to defect recombination. At the critical value $\tilde{\beta}^* = 5.4$, a saddle (open circle) in the energy landscape emerges. The solid red and dashed black lines show the unstable and stable invariant manifolds of this saddle. For $\tilde{\beta} > 5.4$, the saddle separates a stable fixed point at non-zero defect separation (black disk) from the defect free state.}
    \label{fig:recombination}
\end{figure}

We prescribe two pairs of $\pm\frac12$ defects, with $+\frac12$ defects at $(\pm d_{++}/2,0)$ and $-\frac12$ defects at $(\pm (d_{++}/2 + d_{+-}),0)$. The corresponding director angle minimizing $\int\!\mathrm{d} x^2 |\nabla \psi|^2$ is given by
\begin{multline} \label{eq:two-pm-pairs}
    \psi = \frac{\pi}{2} + \frac{1}{2} \arctan (x - d_{++}/2, y) + \frac{1}{2} \arctan (x + d_{++}/2, y) \\
    {}-\frac{1}{2} \arctan [x - (d_{+-} + d_{++}/2), y] - \frac{1}{2} \arctan [x + (d_{+-} + d_{++}/2), y].
\end{multline}
The elastic energy of this configuration is given by
\begin{equation}
    E_\mathrm{el}^{(1)} = \frac{\pi K}{4} \Big[
        -\log (d_{++}) - \log (d_{++} + 2 d_{+-}) + 2 \log (d_{+-}) + 2 \log (d_{++} + d_{+-})
    \Big]
\end{equation}
The alignment energy $E_\mathrm{a}^{(1)}$ is obtained by numerically integrating Eq.~\eqref{eq:alignment-integral} as a function of $d_{++}$ and $d_{+-}$.
Fig.~\ref{fig:recombination} shows the resulting energy landscapes for different values of $\tilde{\beta}$. For $\tilde{\beta} < 5.4$, there is only one minimum at $d_{+-} \to 0$ corresponding to the defect-free state where the defect pairs have recombined. For $\tilde{\beta} > 5.4$, a second minimum appears (black disk) at non-zero $d_{+-}$. This minimum, corresponding to a fixed point (node) in the relaxational flow is separated from the defect free state by a saddle (open white circle). At the critical value $\tilde{\beta}^* \approx 5.4$, the node and the saddle annihilate in a saddle-node bifurcation, corresponding to the defect-recombination transition.
Phrased differently, for $\tilde{\beta} > 5.4$ there is an energy barrier separating the defect-laden state from the defect-free state. This energy barrier vanishes in the saddle-node bifurcation at $\tilde{\beta}^* \approx 5.4$.
Saddle-node bifurcations are the hallmark of systems with multistability (and hence hysteresis).

Near the saddle-node bifurcation, the energy barrier that separates the defect-laden state from the defect free state becomes very shallow such that stochastic fluctuations could drive defect recombination. 
Similarly, fluctuations could drive defect unbinding below the unbinding threshold.
However, since the myonemes are supracellular structures, spanning several cell diameters in length, thermal fluctuations unlikely to play a role.
Rather, active fluctuations due to pulsatile acotmyosin contractility could act as an effective source of noise on the cellular scale. Taking such fluctuations into account is beyond the scope of this work as it would require a careful quantification of the elastic properties of the tissue and the amplitude and correlations of myosin-driven active fluctuations.

The critical value for defect recombination estimated using perturbation theory is a factor $\sim 2$ off from the one found in full numerical simulations ($\tilde{\beta}^*_\mathrm{recomb.} \approx 12$); see Fig.~\ref{fig:planar}(b).
In summary, perturbation theory predicts the correct qualitative behavior for $+\frac12$ defect separation and $\pm\frac12$ defect pair recombination, but is not quantitatively accurate. 

\end{document}

%% file: main.bbl
%